\documentclass[aps,prd,preprintnumbers,showpacs,showkeys,twocolumn,nofootinbib]{revtex4-1}
\usepackage{amsmath}
\usepackage[latin1]{inputenc}
\usepackage[]{caption2}
\usepackage{graphics}
\usepackage[demo]{graphicx}
\usepackage{hyperref}
\usepackage{lipsum}
\usepackage{wrapfig, blindtext}
\usepackage{comment}
\usepackage{tabularx}
\usepackage{float}
\usepackage{slashed}
\usepackage{amssymb}

\begin{document}
\title{Exotic properties of $N^*(1895)$ and its impact on photoproduction of light hyperons}
\author{K.~P.~Khemchandani}\email[email:]{kanchan.khemchandani@unifesp.br}\affiliation{
Universidade Federal de S\~ao Paulo, C.P. 01302-907, S\~ao Paulo, Brazil.}

\author{A.~Mart\'inez~Torres}\email[email:]{amartine@if.usp.br}\affiliation{Universidade de Sao Paulo, Instituto de Fisica, C.P. 05389-970, Sao Paulo,  Brazil.}
\author{Sang-Ho Kim}\email[email:]{sangho.kim@ssu.ac.kr, {\bf Author's new address}: Department of Physics and Origin of Matter and Evolution of
Galaxies (OMEG) Institute, Soongsil University, Seoul 06978, Korea}\affiliation{Department of Physics, Pukyong National University (PKNU),
Busan 48513, Korea.}
\author{Seung-il Nam}\email[email:]{sinam@pknu.ac.kr}\affiliation{Department of Physics, Pukyong National University (PKNU), Busan
48513, Korea}
\affiliation{Asia Pacific Center for Theoretical Physics (APCTP), Pohang 37673,
Korea}
\author{ A.~Hosaka$^{5,6}$}\email[email:]{hosaka@rcnp.osaka-u.ac.jp}\affiliation{Research Center for Nuclear Physics (RCNP), Osaka University,
Ibaraki, Osaka, 567-0047, Japan}
\affiliation{Advanced Science Research Center, Japan Atomic Energy Agency
(JAEA), Tokai 319-1195, Japan}
\preprint{}

\date{\today}

\begin{abstract}
In this work, we outline the findings of our recent  study of the properties of $N^*(1895)$ and its consequential impacts  on the cross sections of the photoproduction of $\Lambda(1405)$. Further, we discuss the possibility of the existence of an isovector state, with the mass similar to $\Lambda(1405)$, which we refer to as $\Sigma(1400)$. With the idea of motivating experimental investigations of $\Sigma(1400)$, we have studied its photoproduction process and determined the respective cross sections and polarization observables. We have studied also the coupling of  $N^*(1895)$ to $K\Sigma(1400)$ and found that it gives an important contribution to the cross sections near the threshold.  In the process, we have determined  the branching ratios of the decay of $N^*(1895)$, to the final states involving $\Lambda(1405)$ and $\Sigma(1400)$, to be in the range of 6-7 MeV. Our findings can motivate consideration of alternative processes in partial wave analyses of experimental data, when studying
  the properties of  $N^*(1895)$. 

\end{abstract}

\keywords{Properties of unstable baryons, Effective Lagrangian approach}
\pacs{13.30.-a, 14.20.Gk, 14.20.Jn}
\maketitle

\renewcommand \thesection{\arabic{section}}

\section{Introduction}
The  nature of  interactions in meson-baryon systems with strangeness $-1$ leads to several interesting questions, like the existence of exotic nuclear matter, the kaonic-nuclei, the possibility of the presence of strange matter in neutron stars, etc (see Refs.~\cite{FRASCATI-DAFNE-AMADEUS:2015wts,Skurzok:2018bur,Friedman:2018rgh} for recent discussions on the topic). The key to finding an answer to these questions is understanding the nature of hyperons, especially $\Lambda(1405)$.  Indeed, there exists an extensive discussion on the nature of $\Lambda(1405)$, attributing its origin to the meson-baryon dynamics (some of most cited works on the topic are Refs.~\cite{Jido:2003cb,Hyodo:2011ur,GarciaRecio:2002td,Dalitz:1967fp,Magas:2005vu,Mai:2012dt,Kaiser:1995eg,osetramos,Oller:2000fj,Mueller-Groeling:1990uxr}). The state is, hence, referred to as a dynamically generated hyperon or a molecular hyperon. Such a description is what raises the questions of binding more and more nucleons to kaons (through coupled channel interactions). Furthermore, there are persistent discussions on associating two complex energy poles to $\Lambda(1405)$, one coupling strongly to $\bar K N$ and the other to $\pi \Sigma$ (for a latest review, see Ref.~\cite{Mai:2020ltx}).
 Most theoretical investigations agree on the position of one of the poles, which lies closer to the real axis and to the $\bar K N$ threshold. However, different works differ 
on the position of the pole lying deeper in the complex plane. It is also important to acknowledge the growing efforts made by the experimental~\cite{Moriya:2013eb,Lu:2013nza,Moriya:2013hwg} and lattice groups~\cite{Hall:2014uca,Menadue:2011pd,Ishii:2007ym,Takahashi:2010nj,Hall:2016kou,Gubler:2016viv,Briceno:2017max}, clearly showing the importance of the topic. It is important to mention before continuing the discussions that although our focus is on strangeness $-1$ here, the existence of exotic nuclear matter without strangeness is also a hot topic of research~\cite{Bass:2018xmz,Kelkar:2013lwa}.

Though $\Lambda(1405)$ gets the dominant focus in the strangeness $-1$ sector, some works indicate the possibility of the presence of a similar, but isovector, hyperon~\cite{Oller:2000fj,Guo,Wu:2009tu,Wu:2009nw,Gao:2010hy,Xie:2014zga,Xie:2017xwx,Khemchandani:2012ur} in the experimental data. The situation is far from clear and more studies are required to reach any conclusions. Our study of meson-baryon interactions with  strangeness $-1$, in Ref.~\cite{Khemchandani:2018amu}, showed that the existing experimental data indicate the presence of such an isovector  state. We refer to it as $\Sigma(1400)$. The experimental data considered in Ref.~\cite{Khemchandani:2018amu}, were total cross sections on different processes with $K^-p$ initial state, energy level shift and width of the $1s$ state of the kaonic hydrogen, and the ratios of cross sections of the different processes near the $K^-p$ threshold.

To motivate experimental studies of $\Sigma(1400)$, we determine different observables for its photoproduction~\cite{Kim:2021wov}. At the same time, we study the photoproduction of $\Lambda(1405)$, on which experimental data is already available. In this way we test our model by comparing our results with the experimental data. Also, we calculate polarization observables for both hyperons, on which data is not available. Our results can be useful for future experimental studies, including those on $\Lambda(1405)$. In fact better statistics data on $\Lambda(1405)$ production is expected to come in future from the ELSA facility~\cite{Scheluchin:2020mhn}. 

The model of photoproduction of the light hyperons, considered in our work, is based on different $s$-, $t$-, and $u$-channel diagrams, including the exchange of relevant hyperon/nucleon resonances using effective Lagrangians. The couplings at the different vertices are determined from the coupled channel meson-baryon amplitudes, which generate the light hyperon resonances. To extend the applicability of the model beyond the threshold region, we use the Regge approach. We also consider the exchange of different nucleon resonances with  mass close to the $KH^*(1400)$ threshold, where $H^*(1400)$ stands for $\Lambda(1405)$ and $\Sigma(1400)$. We find that the most important contribution comes from $N^*(1895)$, which was found to have large meson-baryon coupling in Ref.~\cite{Khemchandani:2013nma}. In the mentioned work, two poles with overlapping widths were found to arise in the complex plane, which interfered and produced a single peak on the real axis around 1900 MeV. The decay properties of $N^*(1895)$ were investigated more recently, in Ref.~\cite{Khemchandani:2020exc}, using its coupling to different meson-baryon channels obtained in the earlier work~\cite{Khemchandani:2013nma}. The study showed that the decay width of $N^*(1895)$ to $KH^*$ is considerable, thus, offering alternative processes to study the properties of the nucleon resonance. 

\section{Formalism and results}
Let us begin the discussions by  briefly mentioning the Lagrangians considered in our works to study meson-baryon scattering. For the vector-meson baryon interactions, we write~\cite{Khemchandani:2011et} 
\begin{eqnarray} \label{vbb}
&\mathcal{L}_{\textrm VB}& = -g \Biggl\{ \langle \bar{B} \gamma_\mu \left[ V_8^\mu, B \right] \rangle\!+\! \langle \bar{B} \gamma_\mu B \rangle  \langle  V_8^\mu \rangle  
\Biggr. \\\nonumber
&+& \frac{1}{4 M}\! \left( F \langle \bar{B} \sigma_{\mu\nu} \left[ V_8^{\mu\nu}, B \right] \rangle \! +\! D \langle \bar{B} \sigma_{\mu\nu} \left\{V_8^{\mu\nu}, B \right\} \rangle\right)\\\nonumber
&& +  \Biggl.  \langle \bar{B} \gamma_\mu B \rangle  \langle  V_0^\mu \rangle  
+ \frac{ C_0}{4 M}  \langle \bar{B} \sigma_{\mu\nu}  V_0^{\mu\nu} B  \rangle  \Biggr\},
\end{eqnarray}
where the subscript $8$ ($0$) denotes the octet (singlet) part of the wave function of the vector meson (relevant in the case of $\omega$ and $\phi$), $V^{\mu\nu}$ represents the  tensor field of the vector mesons,
\begin{equation}
V^{\mu\nu} = \partial^{\mu} V^\nu - \partial^{\nu} V^\mu + ig \left[V^\mu, V^\nu \right], \label{tensor}
\end{equation}
and
\begin{align}
g=\frac{m_v}{\sqrt{2}f_v}, \label{ksrf}
\end{align}
with $m_v (f_v)$ being the mass (decay constant) of a given vector meson in the vertex and the constants $D$ = 2.4, $F$ = 0.8, and $C_0 = 3F - D$.

To determine the  amplitudes for the systems consisting of pseudoscalar mesons and  baryons, we use the lowest order chiral Lagrangian
~\cite{Meissner:1993ah,ecker,pich,Kaiser:1995eg,Oller:2000fj,osetramos,Oller:2006yh}
\begin{eqnarray}\nonumber
\mathcal{L}_{PB} &=& \langle \bar B i \gamma^\mu \partial_\mu B  + \bar B i \gamma^\mu[ \Gamma_\mu, B] \rangle - M_{B} \langle \bar B B \rangle\\
&+&  \frac{1}{2} D^\prime \langle \bar B \gamma^\mu \gamma_5 \{ u_\mu, B \} \rangle + \frac{1}{2} F^\prime \langle \bar B \gamma^\mu \gamma_5 [ u_\mu, B ] \rangle,\label{LPB}
\end{eqnarray}
where  $u_\mu = i u^\dagger \partial_\mu U u^\dagger$, and
\begin{eqnarray}
\Gamma_\mu &=& \frac{1}{2} \left( u^\dagger \partial_\mu u + u \partial_\mu u^\dagger  \right), \\%\,u_\mu = i u^\dagger \partial_\mu U u^\dagger ,\,\\
\, U=u^2 = {\textrm exp} \left(i \frac{P}{f_P}\right),\label{gammau}
\end{eqnarray}
 with $f_P$ representing the pseudoscalar decay constant, and $P$ ($B$) denoting the matrices of the octet meson (baryon) fields, and  $F^\prime = 0.46$ and $D^\prime = 0.8$.

The  transition amplitudes between the pseudoscalar-baryon and the vector-baryon channels are deduced through~\cite{pbvb}
\begin{eqnarray} \label{pbvbeq}
\mathcal{L}_{\rm PBVB} &=& \frac{-i g_{PBVB}}{2 f_v} \left( F^\prime \langle \bar{B} \gamma_\mu \gamma_5 \left[ \left[ P, V^\mu \right], B \right] \rangle \right.\\\nonumber &+&
\left. D^\prime \langle \bar{B} \gamma_\mu \gamma_5 \left\{ \left[ P, V^\mu \right], B \right\}  \rangle \right).
\end{eqnarray}

Using the aforementioned Lagrangians we determine the amplitudes for different meson-baryon systems, for different $s$-, $t$-, and $u$-channels and contact interaction (see Refs.~\cite{Khemchandani:2011et,Khemchandani:2013nma,Khemchandani:2018amu} for more details). We solve Bethe-Salpeter equation with these amplitudes  and look for poles in the complex plane, which are identified with baryon resonances. Within such a formalism, dynamical generation of $N^*(1895)$, $\Lambda(1405)$ and $\Sigma(1400)$ was found in the strangeness 0~\cite{Khemchandani:2013nma} and $-1$ systems~\cite{Khemchandani:2018amu}. The complex energy poles related to these states, as found in Refs.~\cite{Khemchandani:2013nma,Khemchandani:2018amu}, are given in Table~\ref{poles}. 
\begin{table}[h!]
\caption{The poles related to $N^*(1895)$, $\Lambda(1405)$ and $\Sigma(1400)$ as obtained in Refs.~\cite{Khemchandani:2013nma,Khemchandani:2018amu}.}\label{poles}
%\begin{ruledtabular}
\begin{tabular}{ccc}
\hline\hline
State&  \multicolumn{2}{c}{Pole position [MeV]}\\
& \multicolumn{2}{c}{ $E- i \Gamma/2$}\\\hline
$N^*(1895)$ & $1801-i96\quad$&$1912-i54$\\
$\Lambda(1405)$&$1385-i124\quad$&$1426-i15$\\
$\Sigma(1400)$&\multicolumn{2}{c}{$1399 - i 36$}\\\hline\hline
\end{tabular}
%\end{ruledtabular}
\end{table}

As can be seen in Table~\ref{poles}, two poles are related to both  $N^*(1895)$ as well  as $\Lambda(1405)$.  The strangeness $-1$ amplitudes were constrained to fit  the data on the energy level shift and width of the $1s$ state of the kaonic hydrogen~\cite{Bazzi:2011zj} and cross section data (summarized in Ref.~\cite{landolt}) on the processes: $K^- p \to K^- p$, $\bar K^0 n$, $\eta \Lambda$, $\pi^0 \Lambda$, $\pi^0 \Sigma^0$, $\pi^\pm \Sigma^\mp$.  In case of  nonstrange systems, the amplitudes reproduced well  the isospin 1/2 and  3/2 $\pi N$ amplitudes extracted from partial wave analysis~\cite{arndt} and the $\pi^- p \to \eta n$  and $\pi^- p \to K^0 \Lambda$  cross sections up to a total energy of about 2 GeV.

Before proceeding further, a few words on the known information related to $N^*(1895)$ must be mentioned. This  $S_{11}$ nucleon resonance appears after $N^*(1535)$ and $N^*(1620)$, and has been listed as  $N^*(2090)$ ($J^\pi=1/2^-$) in the compilations earlier than  2012 of the particle data group (PDG). All $1/2^-$ states found in partial wave analyses of relevant experimental data are catalogued together under the label of $N^*(1895)$, hence its properties are associated with large uncertainties. Indeed nucleon resonances around 1890 MeV are not well understood, since most of the information in this energy region is extracted from the data on reactions producing $K\Lambda$ and $K\Sigma$ final states and different $N^*$'s couple with a similar strength to these channels.  For example, different descriptions have been brought forward  for the peak present around 1900~MeV in the $\gamma p \to K^+ \Lambda$ total cross sections~\cite{Mart:1999ed,Anisovich:2011fc,Mart:2019mtq,Mart:2012fa,MartinezTorres:2009cw}.   Furthermore, different quark models~\cite{Isgur:1978xj,Bijker:1994yr,Hosaka:1997kh,Takayama:1999kc} predict a different mass for the third $S_{11}$ state, with the values lying near 2 GeV. 

\begin{figure}[H]
  \centering
   \includegraphics[width=.9\linewidth]{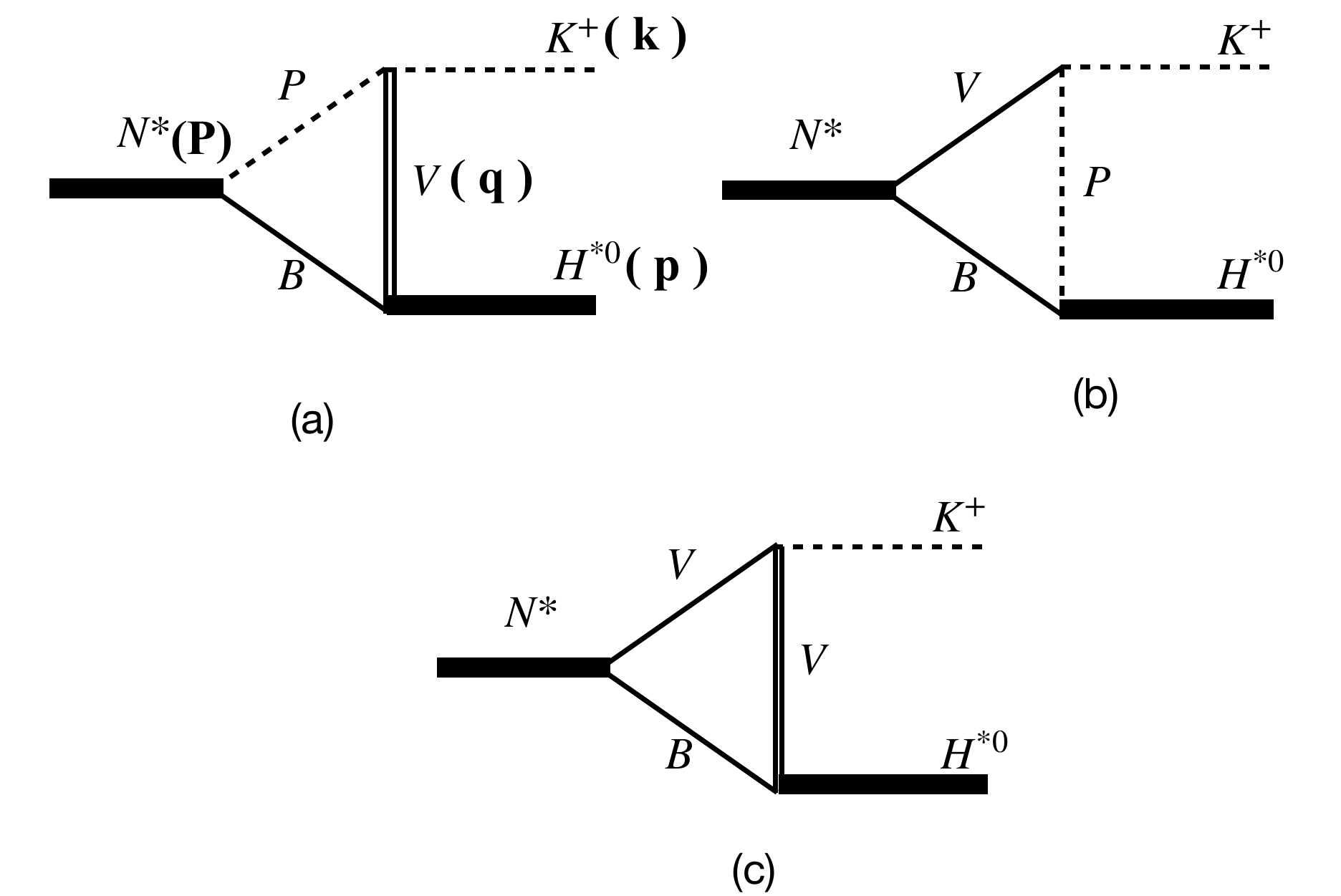}  
    \caption{Different diagrams contributing to the decay of $N^*(1895)$ to $K\Lambda(1405)$ and $K\Sigma(1400)$.}
\label{fig1}
\end{figure}

In such a scenario, detailed studies of the properties of $N^*(1895)$ are very much required. Keeping this in mind, and using the fact that $N^*(1895)$ was found to strongly couple to meson-baryon systems in Ref.~\cite{Khemchandani:2013nma}, we made further investigations of its decay properties in Ref.~\cite{Khemchandani:2020exc}. Using the couplings to different meson-baryon channels  obtained in the previous work~\cite{Khemchandani:2013nma}, we investigated the decay processes of $N^*(1895)$ to final states with $H^*(1400)$ through the diagrams shown in Fig.~\ref{fig1}.
The vertices involving the nucleon/hyperon resonances are written in Ref.~\cite{Khemchandani:2020exc} as
\begin{align}
&\mathcal{L}_{N^*PB}=i g_{PBN^*} \bar B N^* P^\dagger , \nonumber\\
&\mathcal{L}_{N^*VB}=-i\frac{g_{VBN^*}}{\sqrt{3}} \bar B \gamma_5 \gamma_\mu N^* V^{\mu^\dagger} , \nonumber\\
&\mathcal{L}_{PBH^*}=g_{PBH^*} P \bar H^* B , \nonumber\\
&\mathcal{L}_{VBH^*}=i\frac{g_{VBH^*}}{\sqrt{3}} V^{\mu} \bar H^*  \gamma_\mu \gamma_5 B,\label{lageff}
\end{align}
and the remaining ones as
\begin{align}
&\mathcal{L}_{PPV}=-i g_{PPV}\langle V^\mu \left[ P, \partial_\mu P \right] \rangle, \label{vpp} \\
&\mathcal{L}_{VVP}=\frac{g_{VVP}}{\sqrt{2}}\epsilon^{\mu \nu \alpha \beta} \langle \partial_\mu V^\nu \partial_\alpha V_\beta P \rangle. \label{vvp}
\end{align}
The field $H^*$ in  the above equations represents $\Sigma(1400)$ or $\Lambda(1405)$, and the factor $\sqrt{3}$ in the Lagrangians, $\mathcal{L}_{N^*VB}$ and $\mathcal{L}_{VBH^*}$, appears due to the fact that the spin-projected amplitudes were parameterized as Breit-Wigner in Refs.~\cite{Khemchandani:2013nma,Khemchandani:2018amu} when calculating the meson-baryon-resonance couplings.  

The formalism lead to finding of amplitudes (for Fig.~\ref{fig1}(a)) as
\begin{align}
t_a = &i  \sum_j g_{VBH^*\!, j}~ g_{PBN^*\!, j} ~g_{PPV} ~C_j ~\bar u_{H^*}\left(p\right)\gamma_\nu \gamma_5 \nonumber
\\&\int\frac{d^4q}{(2\pi)^4}\Biggl\{ \frac{\left(\slashed{P}-\slashed{k}+\slashed{q}+m_{Bj}\right)}{\left(P-k+q\right)^2-m^2_{Bj}+i\epsilon}\Biggr.\nonumber\\
&\times\left.\frac{\left(-g^{\nu\mu}+\dfrac{q^\nu q^\mu}{m^2_{Vj}}\right)}{q^2-m^2_{Vj}+i\epsilon}\frac{\left(2k -q\right)_\mu}{\left(k-q\right)^2-m^2_{Pj}+i\epsilon}\right\} u_{N^*}\left(P\right),\label{ta1}
\end{align}
where the momenta associated to the different hadrons are as shown in  Fig.~\ref{fig1}(a). The labels $P$, $V$,  and $B$ in Fig.~\ref{fig1} refer to all different possible pseudoscalar, vector meson and baryon channels which can contribute to the loop. All such contributions are summed through the index $j$ in Eq.~(\ref{ta1}).  The constant $C_j$ in Eq.~(\ref{ta1}) is a coefficient coming from  the trace in Eq.~(\ref{vpp}) for the VPP vertex  and $m_{Bj}$, $m_{Vj}$, and $m_{Pj}$ are the masses of the baryon, vector and pseudoscalar meson, respectively, corresponding to the $j$th channel in the  triangular loop. The  values of the different $C_j$ coefficients are  given in Ref.~\cite{Khemchandani:2020exc}.

The product of the spinors, gamma matrices and the numerator of the expression within the curly brackets in  Eq.~(\ref{ta1}) gives the expression
\begin{align}\nonumber
N_a\left(q\right)=&\left(4 ~k\cdot p -2 ~p\cdot q-q^2\right)\bar u_{H^*}\!\left(p\right)\gamma_5u_{N^*}\!\left(P\right)\\
&-2\left(M_{H^*}+m_{Bj}\right)\bar u_{H^*}\!\left(p\right)\slashed k \gamma_5u_{N^*}\!\left(P\right)\nonumber\\
&\left(M_{H^*}+m_{Bj}\right)\bar u_{H^*}\!\left(p\right)\slashed q \gamma_5u_{N^*}\!\left(P\right)\nonumber\\
&+2~\bar u_{H^*}\!\left(p\right)\slashed k\slashed q \gamma_5u_{N^*}\!\left(P\right)+\left(\frac{2 ~k\cdot q-q^2}{m^2_{Vj}}\right)\nonumber\\
&\times\Bigl[\left(M_{H^*}+m_{Bj}\right)\bar u_{H^*}\!\left(p\right)\slashed q \gamma_5u_{N^*}\!\left(P\right)\Bigr.\nonumber\\
&\Bigl.-\left(2~p \cdot q+q^2\right)\bar u_{H^*}\!\left(p\right) \gamma_5u_{N^*}\!\left(P\right)\Bigr],\label{Na}
\end{align}
where $M_{H^*}$ is the mass of $H^*(1400)$.  The integration on the zeroth component of the four momentum in Eq.~(\ref{ta1}) can be done analytically, using Cauchy's theorem. To do so, we reorganize Eq.~(\ref{Na}) by writing it as a series of  terms depending on different powers of $q^0$. In this way, the amplitude gets written as
\begin{align}
t_a &= i \sum_j g_{VBH^*\!,j}~g_{PBN^*\!,j} ~g_{PPV}\mathcal{N}_{H^*}\mathcal{N}_{N^*}C_j \int\frac{d^4q}{(2\pi)^4}\nonumber\\
& \Biggl\{\chi^\dagger\Bigl(\sum\limits_{i=0}^4\mathcal{A}_{i,j}[q^0]^{i}\Bigr)\chi\!\Biggr\} \frac{1}{\left[\left(P-k+q\right)^2-m^2_{Bj}+i\epsilon\right]}\nonumber\\
&\quad\times\!\frac{1}{\!\left[q^2-m^2_{Vj}+i\epsilon\right]\!\left[\left(k-q\right)^2-m^2_{Pj}+i\epsilon\right]}\!,\label{tabis}
\end{align}
where a typical coefficient of $[q^0]^{i}$, for $i=0$, looks like 
\begin{align}
\mathcal{A}_{0,j}&=\vec \sigma\cdot\vec{k}\Biggl\{2\left(M_{H^*}+m_{Bj}\right)+\frac{1}{E_{H^*}+M_{H^*}}\nonumber\\
&\Biggl[\Biggr.2k^0\left(M_{H^*}+m_{Bj}+2E_{H^*} \right)-2\vec k\cdot\vec q+|~\vec q~ |^2+4 |~\vec k~|^2\Biggr.\nonumber\\
&+\left.\left.\frac{|\,\vec q\,|^4+4\left(\vec k\cdot\vec q\right)^2-4\left(\vec k\cdot\vec q\right)|\,\vec q\,|^2}{m^2_{Vj}}\right]\right\}\nonumber\\
&-\vec \sigma\cdot \vec q\left\{ \left(M_{H^*}+m_{Bj}\right)\left(1-\frac{2\vec k\cdot\vec q-|\,\vec q\,|^2}{m^2_{Vj}} \right)\right.\nonumber\\
&+2k^0+\left.2\frac{|~\vec k~|^2}{E_{H^*}+M_{H^*}}\right\}.
\end{align}
The coefficients of other powers of $q^0$ can be found in Ref.~\cite{Khemchandani:2020exc}. 
After integrating on $q^0$, we find the expression to be numerically integrated
\begin{align}
t_a =& i \sum_j g_{VBH^*\!,j}~ g_{PBN^*\!,j} ~g_{PPV}~C_j \mathcal{N}_{H^*}\mathcal{N}_{N^*}\nonumber\\
&\int d\Omega_q \int\limits_{0}^{\Lambda} \frac{d|\,\vec q\,|}{(2\pi)^3} |\,\vec q\,|^2 \sum_{i=0}^{4}\chi^\dagger\Bigl[\mathcal{A}_{i,j}(\,\vec {q}\,)\Bigr]\chi\nonumber\\
&\times \left(\frac{-i N_{i,j}(\,\vec {q}\,)}{\mathcal{D}_j(\,\vec {q}\,)}\right),\label{ta2}
\end{align}
with 
\begin{align}
\frac{-iN_{i,j}(\,\vec {q}\,)}{\mathcal{D}_j(\,\vec {q}\,)}&\equiv\int\frac{dq^0}{\left(2\pi\right)} \frac{\left(q^0\right)^i}{\left[\left(P-k+q\right)^2-m^2_{Bj}+i\epsilon\right]}\nonumber\\
&\frac{1}{\left[q^2-m^2_{Vj}+i\epsilon\right]\left[\left(k-q\right)^2-m^2_{Pj}+i\epsilon\right]}.\label{nbyd}
\end{align}
Detailed analytical expressions for $N_{i,j}$ and $D_j$ are given Ref.~\cite{Khemchandani:2020exc}. A cut-off $\Lambda\simeq 600-700$ MeV is used in the integration on the three-momentum  consistently with the work in Refs.~\cite{Khemchandani:2018amu,Khemchandani:2013nma}, where  relevant experimental data were reproduced by different meson-baryon amplitudes.

Finally, the final states are projected on $p$-wave through
\begin{align}
\langle~\mid t_a\mid~\rangle&=i\! \sum_j\! g_{VBH^*\!,j}~ g_{PBN^*\!,j} ~g_{PPV}\mathcal{N}_{H^*}\mathcal{N}_{N^*}~C_j\nonumber\\
&\times\Biggl\{\frac{1}{2}\int\limits_{-1}^1 d\mathrm{cos}\theta~(-\mathrm{cos}\theta) \int\! d\Omega_q \int\limits_{0}^{\Lambda}\!\frac{d|\,\vec q\,|}{(2\pi)^3} |\,\vec q\,|^2\nonumber\\
& \times\sum_{i=0}^{4}\chi^\dagger_\uparrow\Bigl[\mathcal{A}_{i,j}(~\vec {q}~)\Bigr]\chi_\uparrow\left(\frac{-i N_{i,j}(~\vec {q}~)}{\mathcal{D}_j(~\vec {q}~)}\right)\nonumber\\
&+\frac{1}{2}\int\limits_{-1}^1 d\mathrm{cos}\theta~(-\mathrm{sin}\theta) \int\! d\Omega_q \int\limits_{0}^{\Lambda}\!\frac{d|\,\vec q\,|}{(2\pi)^3} |\,\vec q\,|^2\nonumber\\
& \times \sum_{i=0}^{4}\chi^\dagger_\downarrow\Bigl[\mathcal{A}_{i,j}(~\vec {q}~)\Bigr]\chi_\uparrow\left(\frac{-i N_{i,j}(~\vec {q}~)}{\mathcal{D}_j(~\vec {q}~)}\right)\Biggr\},\label{tapw2}
\end{align}
where the directions of the arrows in the subscript indicate the spinors of $H^*$ and $N^*$.

The amplitudes for other diagrams in Fig.~\ref{fig1} are obtained similarly and explicit expressions can be found in Ref.~\cite{Khemchandani:2020exc}. The results obtained on the branching ratios are summarized in Table~\ref{Br}, where the subscripts on $\Lambda^*$ and $N^*(1895)$ refer to the two poles related to each one of them.  
 \begin{table}[h!]
\caption{Branching ratios (in the isospin base) of the two poles of $N^*(1895)$ to different final states.}\label{Br}
\begin{tabular}{ccccc}
\hline
Decay channel ~~~~& \multicolumn{2}{c}{ Branching ratios ($\%$)}  & ~~~~Experimental   \\
& $N_1^*(1895)$ & $N_2^*(1895)$&data~\cite{pdg} \\\hline \hline
$\pi N$             &9.4    &  10.8     &   2-18  \\
$\eta N$           &2.7    &   18.1    &   15-40\\
$K \Lambda$   &10.9  &  19.4     &   13-23 \\
$K \Sigma$      &0.7    &   26.0    &    6-20\\
$\rho N$           &5.6    &  3.5     &     $<$18\\
$\omega N$     &25.7  &   6.2    &       16-40\\
$\phi N$             &8.9   &   1.1     &      --\\
$K^* \Lambda$ &12.1 &  14.0     &      4-9 \\
$K^* \Sigma$    &6.1   &   0.3    &      -- \\
$K \Lambda_1^*$& 5.4 $\pm$ 0.7& 1.8$\pm$ 0.1&--\\
$K \Lambda_2^*$& 3.3 $\pm$ 0.4& 1.1$\pm$ 0.2&--\\
$K \Sigma^*$& 5.9 $\pm$ 0.8& 11.2$\pm$ 1.1&--\\\hline
\end{tabular}
\end{table}
It can be seen that the results obtained are in good agreement with the available data and that the branching ratios for decay to light hyperons are comparable to meson-baryon decay channels. It should be recalled at this point that  no free parameters are involved in the determination of the partial decay widths. The couplings for each vertex have been taken from our former works~\cite{Khemchandani:2018amu,Khemchandani:2013nma}, where $N^*(1895)$ and $H^*(1400)$ have been found to arise from meson baryon dynamics. Such information can be helpful in using experimental data on $\Lambda(1405)$ production to study the properties of $N^*(1895)$.

\begin{figure}[H]
    \centering
  \includegraphics[width=.9\linewidth]{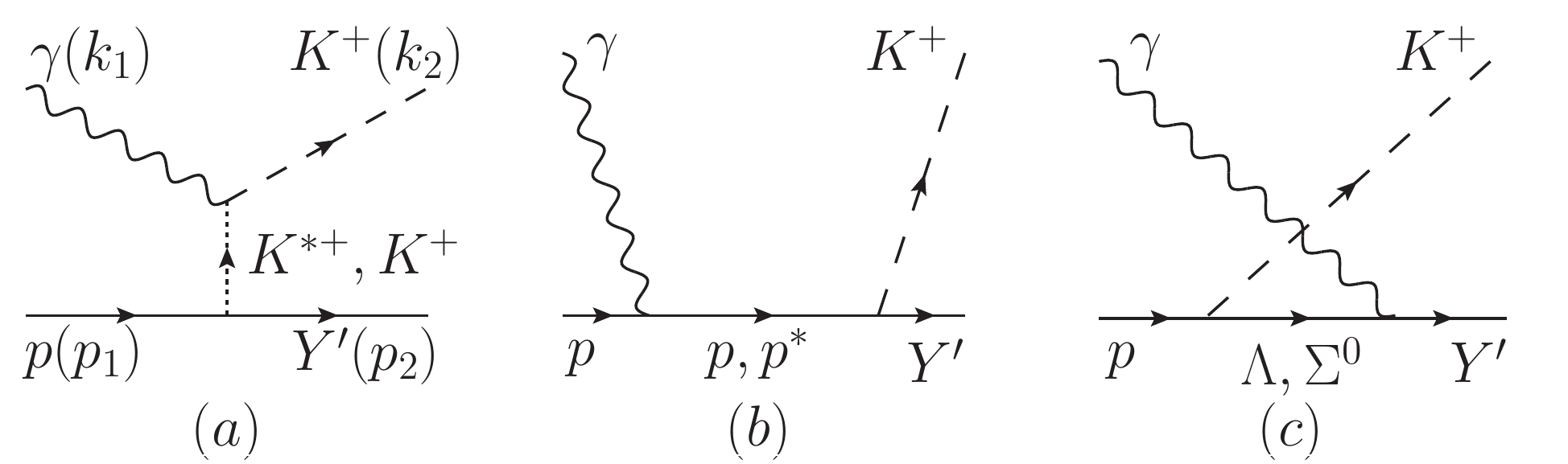}  
    \caption{Diagrams contributing to  the photoproduction of $\Lambda(1405)$ and $\Sigma(1400)$ (denoted as $Y^\prime$ in the diagrams).}
\label{fig2}
\end{figure}

Our findings summarized in Table~\ref{Br}  imply that the exchange of $N^*(1895)$ in the s-channel can play an important role in describing processes with a $K\Lambda(1405)$ final state. In fact we studied the photoproduction of $\Lambda(1405)$ in Ref.~\cite{Kim:2021wov}, considering the  diagrams shown in Fig.~\ref{fig2}
As can be seen from Fig.~\ref{fig2}, we consider the exchange of (several) nucleon resonances (those lying close to the $KH^*(1400)$ threshold). To consider the exchange of $N^*(1895)$ we determined amplitudes for $N^*(1895)\to KH^*(1400)$ vertex, where the mass of $N^*(1895)$ and $H^*(1400)$ vary within the respective range allowed by the respective, associated, widths. Such amplitudes can be found in Ref.~\cite{Khemchandani:2020exc}. 

Further, we determined the radiative decays of $N^*(1895)$ and $H^*(1400)$, to calculate the diagrams shown in Fig.~\ref{fig2}. To do this, we consider the vector meson dominance mechanism and use the Lagrangian~\cite{Roca:2003uk}
\begin{align}
\mathcal{L}_{V\gamma} =-\frac{e F_V}{2} \lambda_{V\gamma} V_{\mu \nu} A^{\mu \nu},\label{vgamma}
\end{align}
where $F_V$ is the decay constant for vector mesons, 
\begin{align}\nonumber
A^{\mu \nu}=\partial^\mu A^\nu-\partial^\nu A^\mu,
\end{align}
 $V_{\mu \nu}$ is a tensor field related to $\rho^0$, $\omega$, anf $\phi$, with $\lambda_ {V\gamma} = 1,~\frac{1}{3}$ and $-\frac{\sqrt{2}}{3}$, respectively, and 
\begin{align}
V^{\mu \nu}= \frac{1}{M_V}\left(\partial^\mu V^\nu-\partial^\nu V^\mu\right).
\end{align}
Consequently we obtain the amplitude for the $B^* \to B \gamma$ process  as
\begin{align}
t_{B^*\to B \gamma} = \frac{2 e F_V \tilde g_{VBB^*} \lambda_{V\gamma}}{ M_V^2} \bar B \gamma_5 \slashed \epsilon \slashed K B^*,
\end{align}
with $\epsilon$ denoting the polarization vector for the photon and plug it in the following equation to determine the width,
 \begin{align}
 \Gamma_{B^* \to B \gamma} &= \frac{1}{32 \pi^2}\frac{|~\vec K~|\left( 4 M_{B^*} M_{B}\right)}{M_{B^*}^2}\frac{1}{2 S_{B^*} +1}\nonumber\\&\times \int d\Omega \sum_{m_{B^*}, m_{B}, m_\gamma} |\mathcal{M}_{B^* \to B \gamma}|^2. \label{eq:width2}
 \end{align}

The results obtained on the radiative decay widths are given in Table~\ref{TAB3}.
\begin{table}[H]
\caption{Radiative decay widths for $\Lambda(1405)$, $\Sigma(1400)$ and $N^*(1895)$. In all the results shown here, an interference between the two poles related to the decaying hadron has been considered to obtain the decay width.}
\label{TAB3}\vspace{0.3cm}
\centering
\begin{tabular}{cc}
\hline\hline
Decay process& Partial width [KeV] \\ 
\hline
$\Lambda(1405) \to \Lambda \gamma$&26.19 $\pm$ 6.93\\
$N^*(1895) \to p \gamma$ &650.70 $\pm$ 65.10\\
$\Sigma(1400) \to \Lambda \gamma$&49.97 $\pm$ 8.57\\
$\Sigma(1400) \to \Sigma \gamma$&94.51$ \pm$ 9.33\\
$\Lambda(1405) \to \Sigma \gamma$&2.50 $\pm$ 1.37\\
\hline\hline
\end{tabular}
\end{table}
The radiative decay width of $\Lambda(1405)\to\Lambda \gamma$, listed by PDG,  is  $27\pm8$~KeV~\cite{pdg}. It can be seen that  our result is in remarkable agreement with the experimental data~\cite{pdg}. For $\Lambda(1405)\to\Sigma \gamma$, PDG~\cite{pdg} provides two possible values: $10\pm4$~KeV or $23\pm7$~KeV. Our results are closer to the former value. It is more complex to draw conclusions on the partial width obtained for the $N^*(1895)$ radiative decay by comparing them with the information provided by PDG, since two complex energy poles associated with it will be required in partial wave analysis. Still, we can say that our results for the second pole seem to be closer  to the upper limit of the value,  0.01-0.06, given in Ref.~\cite{pdg}. The former finding is consistent with a better agreement between the real and imaginary part  of the second pole of $N^*(1895)$ found in our work and the values associated with $N^*(1895)$ in Ref.~\cite{pdg}.  

We are now in a position to discuss the effective Lagrangians used  in our model to calculate the photoproduction cross sections. For the electromagnetic vertices, we write
%EQUATION>>>
\begin{align}
\mathcal L_{\gamma K K} &=
-ie [  K^\dagger (\partial_\mu K) - (\partial_\mu K^\dagger) K ] A^\mu ,       \cr
\mathcal L_{\gamma K K^*} &=
g_{\gamma K K^*}^c \epsilon^{\mu\nu\alpha\beta} \partial_\mu A_\nu
[ (\partial_\alpha K_\beta^{*-}) K^+ + K^- (\partial_\alpha K_\beta^{*+} )] ,    \cr
\mathcal L_{\gamma NN} &=
- e \bar N \left[ \gamma_\mu \frac{1+\tau_3}{2} - \frac{\kappa_N}{2M_N}
\sigma_{\mu\nu}\partial^\nu \right] A^\mu N ,                                \cr
\mathcal L_{\gamma Y Y^*} &=
\frac{e\mu_{Y^*Y}}{2M_N} \bar Y \gamma_5 \sigma_{\mu\nu} \partial^\nu
A^\mu Y^* + \mathrm{H.c.},
\label{eq:Lag:EM}
\end{align}
where $A^\mu$ is the photon field,  $Y$ denotes  $\Lambda (1116)$ or $\Sigma^0(1192)$,
the coupling constant $g_{\gamma K K^*}^c$ is determined from the experimental data
for $\Gamma_{K^{*+} \to K^+ \gamma}$~\cite{pdg}, to be 0.254
$\mathrm{GeV}^{-1}$,  $\kappa_p$ = 1.79~\cite{pdg} is the proton anomalous magnetic moment, and $\mu_{Y^*Y}$ represents transition magnetic moments.

Further, effective Lagrangians required at the strong vertices are taken as
%EQUATION>>>
\begin{align}
\mathcal L_{K N Y} &=
- i g_{K N Y} \bar N \gamma_5 Y K + \mathrm{H.c.},                           \cr
%\mathcal L_{K N Y} &=
%\frac{g_{K N Y}}{M_N+M_Y} \bar N \gamma_\mu \gamma_5 Y \partial^\mu K +
%\mathrm{H.c.},                                                            \cr
\mathcal L_{K N Y^*} &=
%g_{K N Y^*} \bar N Y^* K + \mathrm{H.c.},                                  \cr
g_{K N Y^*} \bar K \bar Y^* N + \mathrm{H.c.},                                \cr
\mathcal L_{K^* N Y^*} &=
%-i\frac{g_{K^* N Y^*}}{\sqrt3} \bar N \gamma_\mu \gamma_5 Y^* K^{*\mu}
i\frac{g_{K^* N Y^*}}{\sqrt3} \bar K^{*\mu} \bar Y^* \gamma_\mu \gamma_5 N
+ \mathrm{H.c.},
\label{eq:Lag:Strong}
\end{align}
where  $g_{K N(\Lambda,\,\Sigma^0)} =(-13.4,\,4.09)$~\cite{Stoks:1999bz,Rijken:1998yy}.
The values for the $Y^*\bar K N$ and $Y^*\bar K^*N$ couplings  are taken from Ref.~\cite{Khemchandani:2018amu}. The factor $1/\sqrt{3}$ in $\mathcal{L}_{K^*NY^*}$ takes into account the fact that the Lagrangian has a spin structure while the couplings in
 Refs.~\cite{Khemchandani:2018amu,Khemchandani:2013nma} are obtained by evaluating  the residue of the spin projected $t$-matrices.
 
In order to extend the applicability of our model beyond the threshold region we use a hybridized Regge model.
This is done by replacing the Feynman propagators for $K$ and $K^*$  by the Regge expressions
%EQUATION>>>
\begin{align}\nonumber
&\frac{1}{t-M_K^2}  \to  \left( \frac{s}{s_0^K} \right)^{\alpha_K(t)}
\frac{\pi\alpha_K'}{\sin[\pi\alpha_K(t)]}
\left\{ \begin{array}{c} 1 \\ e^{-i\pi\alpha_K(t)} \end{array} \right\}\\
& \times\frac{1}{\Gamma[1+\alpha_K(t)]}       \\       \nonumber                        
& \frac{1}{t-M_{K^*}^2}  \to \left( \frac{s}{s_0^{K^*}} \right)^{\alpha_{K^*}(t)-1}
\frac{\pi\alpha'_{K^*}}{\sin[\pi\alpha_{K^*}(t)]}
\left\{ \begin{array}{c} 1 \\ e^{-i\pi\alpha_{K^*}(t)} \end{array} \right\}\\\nonumber
&\frac{1}{\Gamma[\alpha_{K^*}(t)]},
\end{align}

where
\begin{align}
\alpha_K(t)&= 0.7\, \mathrm{GeV^{-2}} (t - M_K^2)\\\nonumber
\alpha_{K^*}(t)&= 1 + 0.83\, \mathrm{GeV^{-2}} (t - M_{K^*}^2),
\label{eq:ReggeTraj}
\end{align}
with 
\begin{align}
\alpha'_{K,K^*} \equiv \partial \alpha_{K,K^*}(t)/\partial t,
\end{align}
and
$s_0^K = 3.0\, \mathrm{GeV^2}$ and  $s_0^{K^*}= 1.5 \, \mathrm{GeV^2}$.
More details on the formalism can be found in Ref.~\cite{Kim:2021wov}.

\begin{figure}[H]
    \centering
    \includegraphics[width=.8\linewidth]{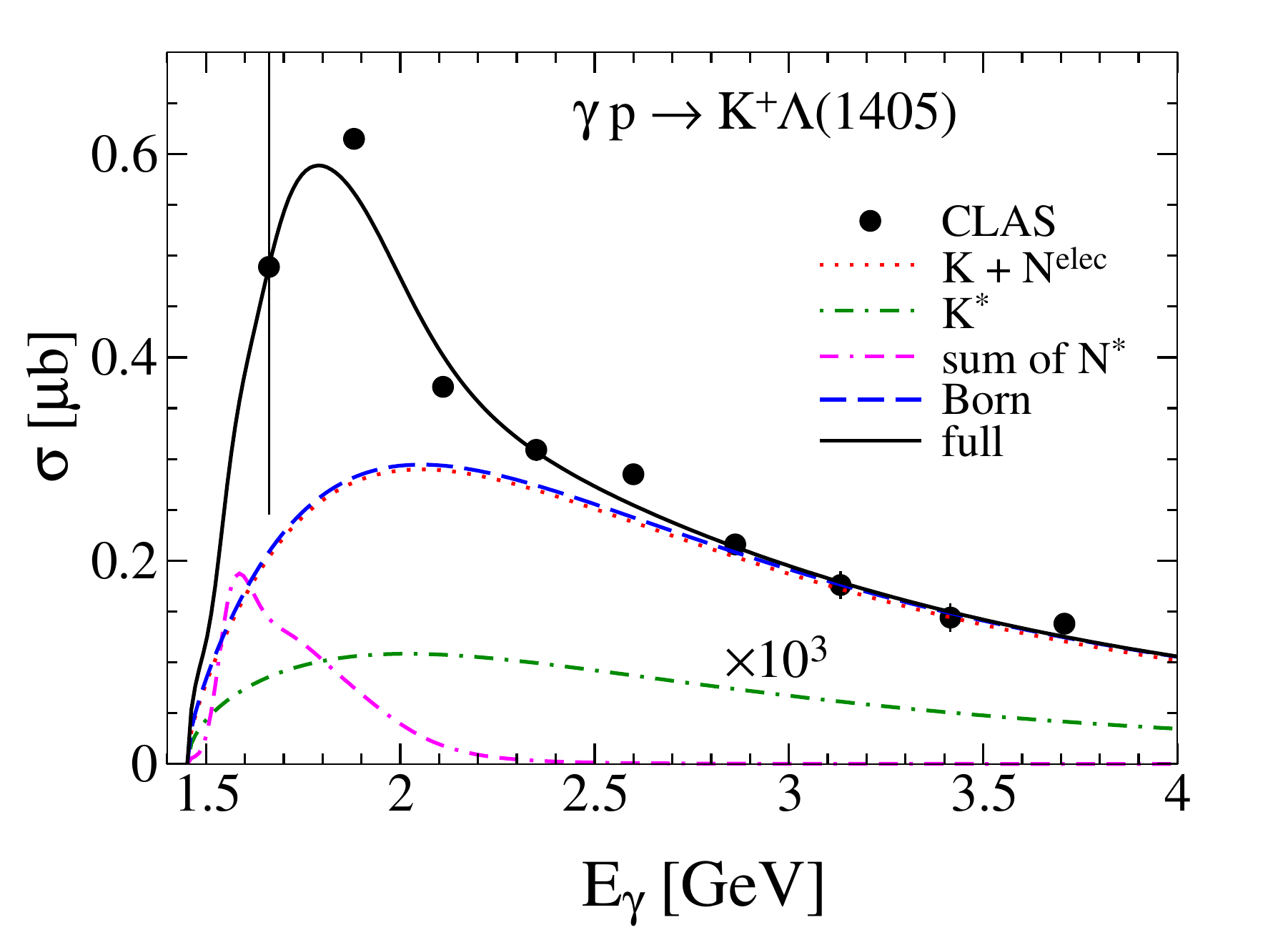}  
 \caption{Total cross section for $\gamma p \to K^+ \Lambda(1405)$  
as a function of the beam energy, $E_\gamma$.
The  dot-double-dashed curve depicts, among other $N^*$'s, the
contribution from the $N^*(1895)$ exchange, which is found to be the dominant.
The data are taken from  Ref.~\cite{Moriya:2013hwg}.
}
\label{XnLambda}
\end{figure}

 In Fig.~\ref{XnLambda},  we  show  our results on the  photoproduction cross sections for the case of $\Lambda(1405)$ as a function of the beam energy. It can be seen that our results (shown as a continuous line) are in good agreement with the data obtained by the  CLAS Collaboration~\cite{Moriya:2013hwg}.  The contributions to the full results from different sources are also shown in Fig.~\ref{XnLambda}. The  dotted curve shows the contribution from  the kaon Reggeon plus the electric part of the nucleon exchange (the sum is a gauge invariant contribution). The  dot-dashed and the  dot-dashed-dashed curves show the $K^*$-Reggeon and the exchange of $N^*$'s, respectively. The  dashed curve depicts the total contributions from the Born diagrams.
The nucleon resonances considered in the s-channel exchange are: $N^*(1895)$, $N^*(2000,\,5/2^+)$ $N^*(2100,\,1/2^+)$, 
$N^*(2030,1/2^-)$, $N^*(2055,\,3/2^-)$, and $N^*(2095,\,3/2^-)$. However, the contributions from the states other than $N^*(1895)$ are found to be small and the exchange of $N^*(1895)$ plays an important role in describing the data in the low-energy region ($E_\gamma \leqslant 2.5$ GeV).

\begin{figure}[H]
   \centering
  \includegraphics[width=.8\linewidth]{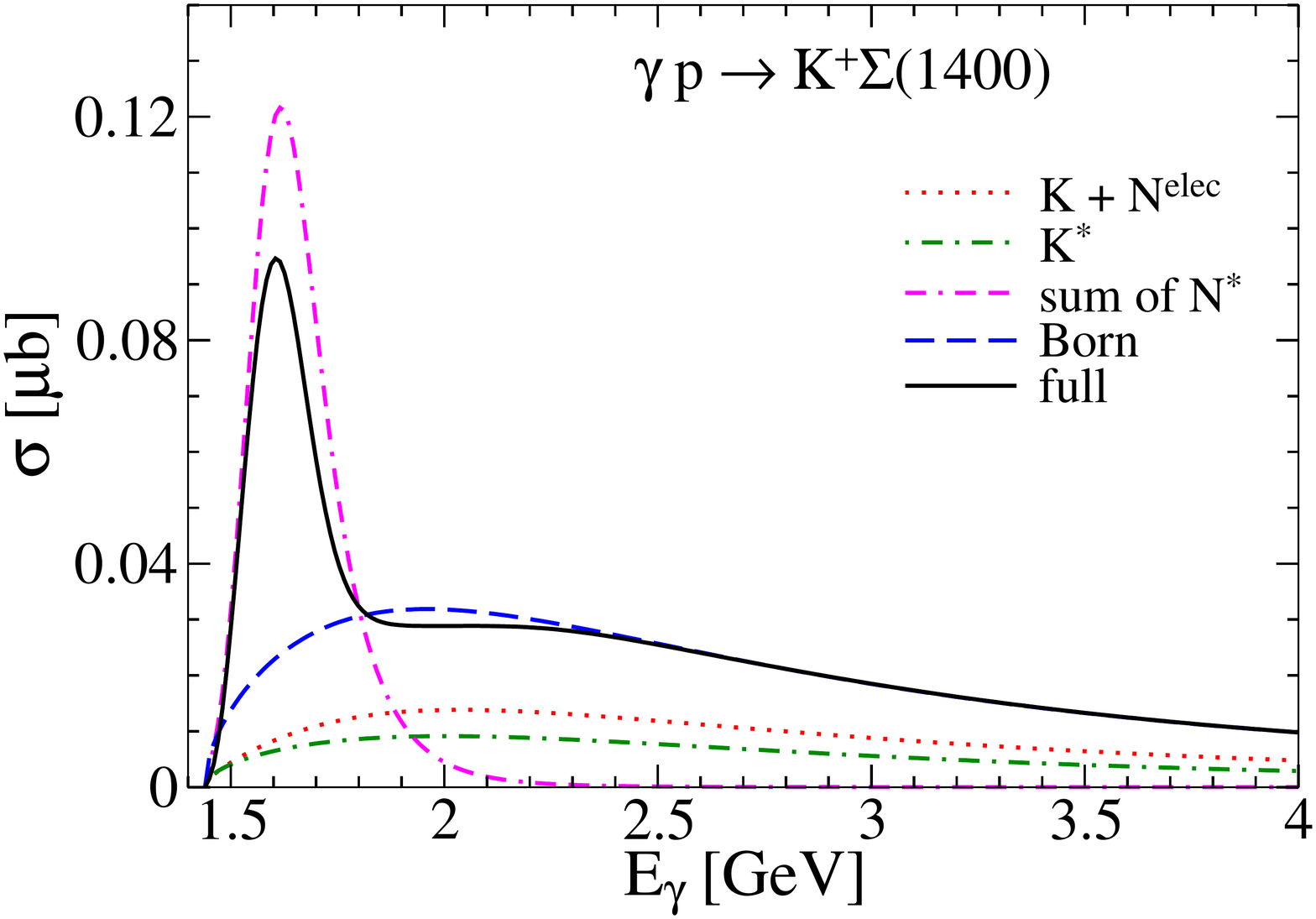}  
    \caption{Total cross section for $\gamma p \to K^+ \Sigma(1400)$ 
as a function of the beam energy, $E_\gamma$.
Here  the (magenta) dot-double-dashed curve depicts the contribution from the s-channel diagram with the $N^*(1895)$ exchange.}
\label{XnSigma}
\end{figure}

We also predict the cross sections for  the process $\gamma p \to K^+ \Sigma(1400)$. The results are shown in Fig.~\ref{XnSigma}. The order of magnitude of the  cross sections  shown in  Fig.~\ref{XnSigma} is measurable at currently available facilities.
 It should be pointed out that also in this case, the $N^*(1895)$ exchange in the s-channel provides important contributions to the cross sections near the threshold. It can  be noticed  that the cross sections coming from the $N^*$-exchange give rise to cross sections  larger than those obtained from the total amplitude, which indicates a negative interference occurs between different contributions.  We hope that our findings motivate future experimental investigations of the photoproduction of $\Sigma(1400)$. It should be mentioned that results on different polarization observables are also shown and discussed in Ref.~\cite{Kim:2021wov}.

\section{Conclusions}
We can summarize the findings of our works presented at the conference by mentioning that the description of the properties of $N^*(1895)$ necessarily requires important contributions from hadron dynamics. This nucleon resonance couples strongly to light hyperons ($\Lambda(1405)$ and $\Sigma(1400)$) and the related decay widths are comparable to the processes involving  other meson-baryon channels. Here $\Sigma(1400)$ is a $1/2^-$ resonance, whose existence is still not certain and requires further experimental investigations. Our findings show that $N^*(1895)$ exchange gives important contributions to the cross sections of photoproduction of  $\Lambda(1405)$ as well as $\Sigma(1400)$.  There exist data on the photoproduction of $\Lambda(1405)$ and our results are in good agreement with this data. Our findings on the cross sections for $\Sigma(1400)$ are predictions and should serve as a motivation for experimental investigations of the process, which should be useful in establishing the existence of the state with better certainty.

\section{Acknowledgements}
K.P.K and A.M.T gratefully acknowledge the  support from the Funda\c c\~ao de Amparo \`a Pesquisa do Estado de S\~ao Paulo (FAPESP), processos n${}^\circ$ 2019/17149-3 and 2019/16924-3, by the Conselho Nacional de Desenvolvimento Cient\'ifico e Tecnol\'ogico (CNPq), grants n${}^\circ$ 305526/2019-7 and 303945/2019-2. A.M.T also thanks the partial support from mobilidade Santander  (edital PRPG no 11/2019). H.N. is supported in part by Grants-in-Aid for Scientific Research (JP17K05443 (C)). AH is supported in part by Grants-in-Aid for Scientific Research (JP17K05441 (C)) and for Scientific Research on Innovative Areas (No. 18H05407).
S.i.N thanks National Research Foundation of Korea funded by the Ministry of Education, Science and Technology (MSIT) (NRF-2018R1A5A1025563 and 2019R1A2C1005697).
The work of S.H.K was supported by Basic Science Research Program through the National Research Foundation of Korea (NRF) funded by the Ministry of
Education (NRF-2021R1A6A1A0304395 and NRF-2022R1I1A1A01054390).

\end{document}